\documentclass[12pt]{article}
\usepackage[dvips]{graphics}
\usepackage[dvips]{color}
\usepackage{epsfig}
\usepackage{psfig}
\usepackage{graphicx}
\usepackage{nicefrac}
\usepackage{amssymb}
\newcommand{\beq}{\begin{equation}}
\newcommand{\eeq}{\end{equation}}
\newcommand{\beqn}{\begin{eqnarray}}
\newcommand{\eeqn}{\end{eqnarray}}

\newcommand{\tr}{{\rm Tr\,}}

\renewcommand{\vec}[1]{{\bf #1}}

\newcommand{\psibar}{\bar\psi}

\makeatletter
\renewcommand\section{\@startsection {section}{1}{\z@}%
                                   {-5.5ex \@plus -1ex \@minus -.2ex}
                                   {2.3ex \@plus.2ex}%
                                   {\normalfont\large\bfseries}}
\renewcommand\subsection{\@startsection{subsection}{2}{\z@}%
                                     {-3.25ex\@plus -1ex \@minus -.2ex}%
                                     {1.5ex \@plus .2ex}%
                                     {\normalfont\normalsize\bfseries}}
\renewcommand\thesection {\@arabic\c@section}
\renewcommand\thesubsection   {\thesection.\@arabic\c@subsection}
\renewcommand{\@seccntformat}[1]{%
\csname the#1\endcsname.\hspace{1.0em}}
\makeatother

\begin{document}

\begin{titlepage}
\begin{centering}
\vfill

\vspace*{1cm}
{\LARGE\bf Exploration of {\boldmath $SU(N_c)$} gauge theory with \\
many Wilson fermions at strong coupling}

\vspace{0.8cm}

Kei-ichi Nagai,
Georgina Carrillo-Ruiz,
Gergana Koleva,
Randy Lewis

{\em Department of Physics and Astronomy, York University,
Toronto, Canada, M3J 1P3}

\vspace*{0.8cm}

\end{centering}
 
\vspace*{0.4cm}

\noindent
\begin{abstract}
We explore aspects of the phase structure of $SU(2)$ and $SU(3)$ lattice
gauge theories at strong coupling with many flavours $N_f$ of Wilson fermions
in the fundamental representation.
The pseudoscalar meson mass as a function of hopping parameter is observed to
deviate from the expected analytic dependence, at least for sufficiently
large $N_f$.  Implications of this effect are discussed, including the
relevance to recent searches for an infrared fixed point.
\end{abstract}

\vspace*{0.4cm}
\noindent
PACS numbers: 
11.5.15.Ha 

\vspace*{1cm}
 
\vfill

\end{titlepage}

\setcounter{page}{2}
\section{Introduction}

There has been a lot of interest recently in identifying an infrared
fixed point, and the possibility of an associated conformal window,
for $SU(N_c)$
gauge theories as a function of the number of fermion flavours $N_f$\cite{
Iwasaki,
Catterall,
Appelquist,
Shamir,
Deuzeman,
DelDebbio,
Fodor,
Hietanen,
Hasenfratz:2009ea}, motivated by the pioneering work of Banks and
Zaks\cite{Banks:1981nn}.
For a valuable review, see Ref.~\cite{Fleming:2008gy}.
Much of the interest stems from the longstanding suggestion that a (nearly)
conformal gauge theory could play a key role in electroweak symmetry breaking
beyond the standard model of particle physics\cite{Holdom,
Yamawaki:1985zg,
AppelquistWij}.
For fermions in the fundamental representation, the prediction obtained from a
detailed study using the Wilson fermion
action\cite{Iwasaki} differs from subsequent studies that use other
actions\cite{Fleming:2008gy}.
According to Ref.~\cite{DeGrand:2009mt} it is generally believed that
confinement exists in the strong coupling limit for any $N_f$; the
disagreement with data from the Wilson action\cite{Iwasaki} suggests that
the unphysical Wilson term has a dramatic effect.
In the present work, we revisit the strong coupling limit of the Wilson action
with the goal of providing some additional insight into this issue.

An $SU(N_c)$ lattice gauge theory with fundamental Wilson fermions is
expected to have a phase where flavour and parity are spontaneously
broken, known as the Aoki phase\cite{Aoki,Aoki:1995ft}.
The expected phase diagram (for example figure 3 of Ref.~\cite{Aoki:1995ft})
is sketched in the plane spanned by the two relevant parameters: the gauge
coupling and the fermion mass.
The weak coupling part of the diagram shows ``fingers'' of Aoki phase
separated by regions of symmetric phase, but ongoing research by various
groups is refining this picture for specific discretizations and
lattice improvement scenarios\cite{Farchioni:2004us,
Michael:2007vn,
Boucaud:2008xu,
Ilgenfritz:2003gw,
Farchioni:2005hf,
Azcoiti:2008dn,
Sharpe:2008ke}.
The strong coupling part of the original expected phase diagram is simpler:
the Aoki phase exists for quark masses of smaller magnitude and the symmetric
phase exists for quark masses of larger magnitude.  In the extreme strong
coupling limit ($g=\infty$), the phase boundary is predicted to be at a
specific critical value of the hopping parameter,
\begin{equation}\label{eq:standardkappa}
\kappa_c = \frac{1}{4} \,.
\end{equation}
Recall that the hopping parameter $\kappa$ is related to the dimensionless
bare fermion mass $m_0$ as follows:
\begin{equation}
\kappa \equiv \frac{1}{2m_0+8} \,.
\end{equation}
In addition, the dimensionless pion mass in the strong coupling limit is
predicted to be a rather simple function of the hopping
parameter\cite{Aoki},
\begin{equation}\label{eq:pion}
\cosh(m_\pi) = 1 + \frac{(1-16\kappa^2)(1-4\kappa^2)}{8\kappa^2(1-6\kappa^2)}
\,.
\end{equation}
This expression for the pion mass was derived long ago for
quenched configurations (i.e.\ $N_f=0$ inside the
configurations)\cite{Wilson:1975id,Kawamoto:1980fd}, and initial tests of its
validity for $N_f>0$ were presented in some of the direct numerical simulations
within Ref.~\cite{Iwasaki}.
A central result of our work is the observation of systematic
deviations from Eq.~(\ref{eq:pion}) in exploratory
simulations with multiple Wilson fermions at $\beta=0$.

\section{Simulation details}

For this study we use the standard plaquette gauge action and
the standard Wilson fermion action for each of the $N_f$ degenerate fermions,
\begin{eqnarray}
S &=& S_G + \sum_{f=1}^{N_f}S_W^f \,, \\
S_G &=& \frac{\beta}{2}\sum_{x,\mu,\nu}\left(1-\frac{1}{N_c}{\rm Re}\tr
        U_\mu(x)U_\nu(x+\mu)U_\mu^\dagger(x+\nu)U_\nu^\dagger(x)\right) \,, \\
S_W^f &=& \sum_x\bigg[ \psibar^f(x) \psi^f(x)
- \kappa\bigg\{\psibar^f(x)\left(1-\gamma_\mu \right) U_\mu(x) \psi^f(x+\mu)
     \nonumber \\ &&
+ \psibar^f(x+\mu)(1+\gamma_\mu)U^{\dagger}_\mu(x)\psi^f(x)\bigg\}\bigg] \,,
\end{eqnarray}
where $\beta=2N_c/g^2$ with $g$ the gauge coupling,
and where $U_\mu(x)$ is the $SU(N_c)$-valued link variable.
Then the partition function $Z$ is
\begin{eqnarray}
Z &=& \int [d U_\mu(x)] \prod_{f=1}^{N_f}[d \psibar^f(x)][d \psi^f(x)]\exp(-S)
      \nonumber \\
  &=& \int [d U_\mu(x)] \left(\det(D_W^{\dagger}D_W)\right)^{\frac{N_f}{2}}
      \exp(-S_G) \,,
\end{eqnarray}
where $D_W$ is the kernel of the single-fermion action 
$S_W^f=\psibar^f(x)D_W(x,y)\psi^f(y)$.

Simulations are performed with a standard Hybrid Monte Carlo (HMC).
For independent confirmation, simulations for the heavier fermions are repeated
using a Polynomial Hybrid Monte Carlo (PHMC)\footnote{Use of our PHMC for the
lighter fermions would require a polynomial of remarkably high degree.}.
In each Molecular Dynamics (MD) evolution, the number of steps per trajectory
$N_{\rm MD}$ is tuned to beyond $80\%$ acceptance for the Metropolis
test.  
The step size $\Delta\tau$ is defined by $\Delta \tau N_{\rm MD}=1$.

Our simulations focus primarily on $\beta=0$, but some studies at $\beta=2$
will also be reported.  Lattice sizes include $6^2\times12^2$, $8^2\times16^2$,
$12^2\times24^2$ and $12^3\times24$.
With zero temperature simulations in mind,
we use periodic boundary conditions and define the Euclidean time
dimension to be greater than or equal to every spatial dimension.
For each chosen set of parameters,
50 to 100 configurations were collected after thermalization,
with 4 or 5 trajectories between saved configurations.

Our primary observables are the average plaquette $\left<\square\right>$,
the pseudoscalar meson mass $m_\pi$, and the axial-Ward-Takahashi-identity
quark mass $m_q^{\rm AWI}$, all defined in standard fashion:
\begin{eqnarray}
\left<\square\right> &=&
\left<\frac{1}{12N_cV}\sum_{x,\mu,\nu}{\rm ReTr}U_\mu(x)U_\nu(x+\mu)
U_\mu^\dagger(x+\nu)U_\nu^\dagger(x)\right> \,, \\
\left<\sum_{\bf x}P({\bf x},t)P({\bf 0},0)\right> &=& c_\pi e^{-m_\pi t}
+ \sum_nc_ne^{-m_nt} ~~~({\rm with}~m_n>m_\pi~\forall n), \\
m_q^{\rm AWI} &=& \frac{\left<\nabla_4\sum_{\bf x}A_4({\bf x},t)P({\bf 0},0)
\right>}{\left<2\sum_{\bf x}P({\bf x},t)P({\bf 0},0)\right>} \,,
\end{eqnarray}
where $V$ denotes the total number of lattice sites,
$\nabla_4$ is the symmetric temporal derivative operator, and
the flavour non-singlet bilinear operators (with flavour index omitted) are
\begin{eqnarray}
P({\bf x},t) &=& \psibar({\bf x},t)\gamma_5\psi({\bf x},t) \,, \\
A_4({\bf x},t) &=& \psibar({\bf x},t)\gamma_4\gamma_5\psi({\bf x},t) \,.
\end{eqnarray}

\section{{\boldmath $SU(2)$} with many flavours at \boldmath{$\beta=0$}}

\begin{figure}[tbh] 
\hspace{2cm}\includegraphics[width=12cm,trim=0 0 0 0,clip=true]{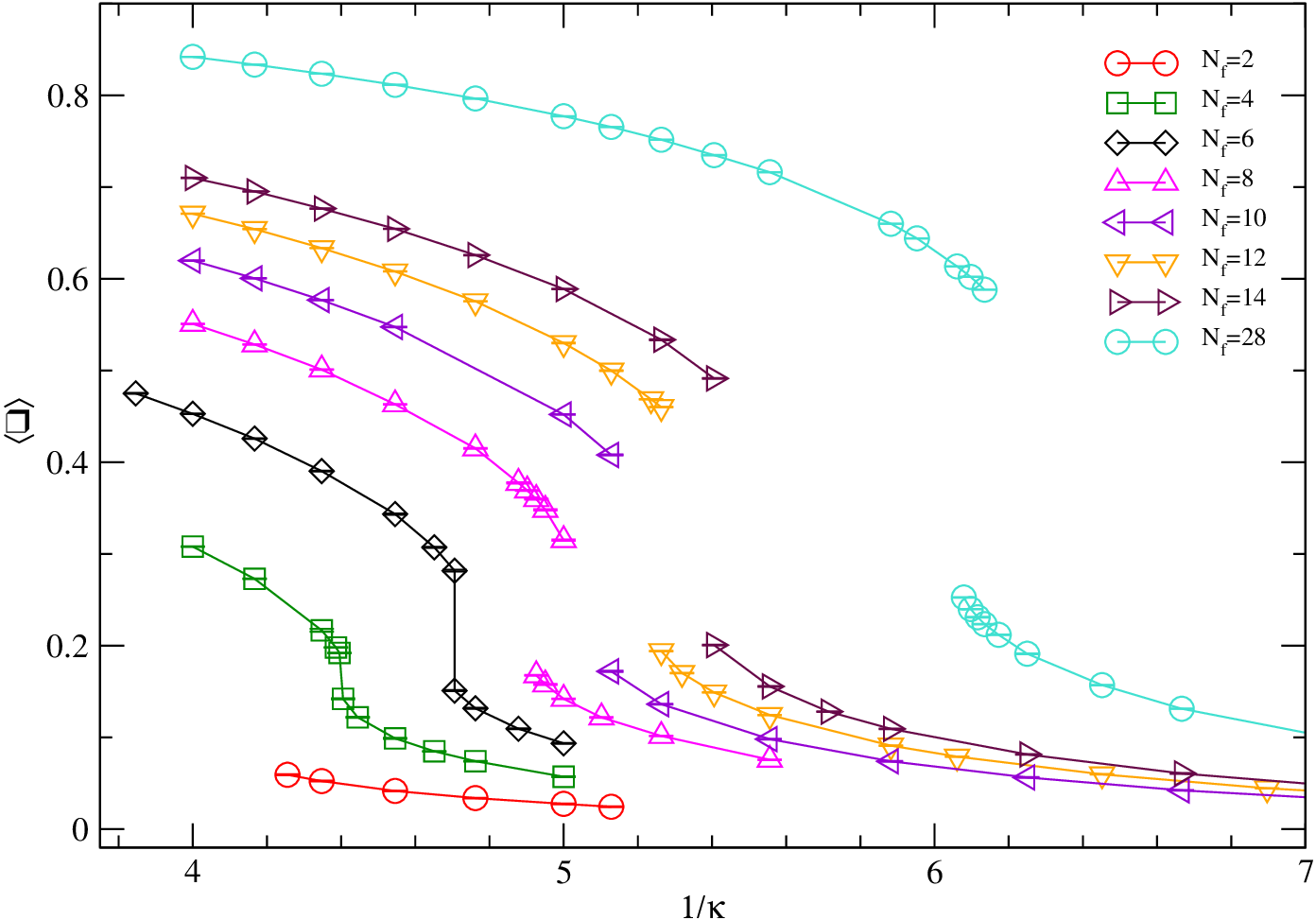}
\caption{
Average plaquette as a function of $1/\kappa$ 
in $SU(2)$ gauge theory with $N_f$ flavours at $\beta=0$.
}
\label{fig:plaqnc2b0}
\end{figure}

For a first indication of the difference 
between quenched and dynamical simulations,
consider the average plaquette for various numbers of flavours, as shown in
Fig.~\ref{fig:plaqnc2b0}.
An abrupt transition is clearly seen for $N_f\geq6$, with a two-state
signal defining the transition region in each case.  The data at $N_f=4$ hint
at a transition but are not conclusive, though other observables to
be discussed below will confirm a $N_f=4$ transition.  At $N_f=2$ the
transition is either suppressed or absent.  This behaviour implies the
existence of a first order transition for sufficiently large $N_f$, and was
already reported for the $SU(3)$
case in Ref.~\cite{Iwasaki} and for $SU(2)$ with two
flavours of adjoint-representation fermions in
Refs.~\cite{Catterall,Hietanen}.
Our results show that the first order transition moves to larger values of
$1/\kappa$ as $N_f$ is increased.
All of our observed transitions are in the range
$\nicefrac{1}{8}<\kappa<\nicefrac{1}{4}$.

\begin{figure}[tbh]
\hspace{2cm}\includegraphics[width=12cm,trim=0 0 0 0,clip=true]{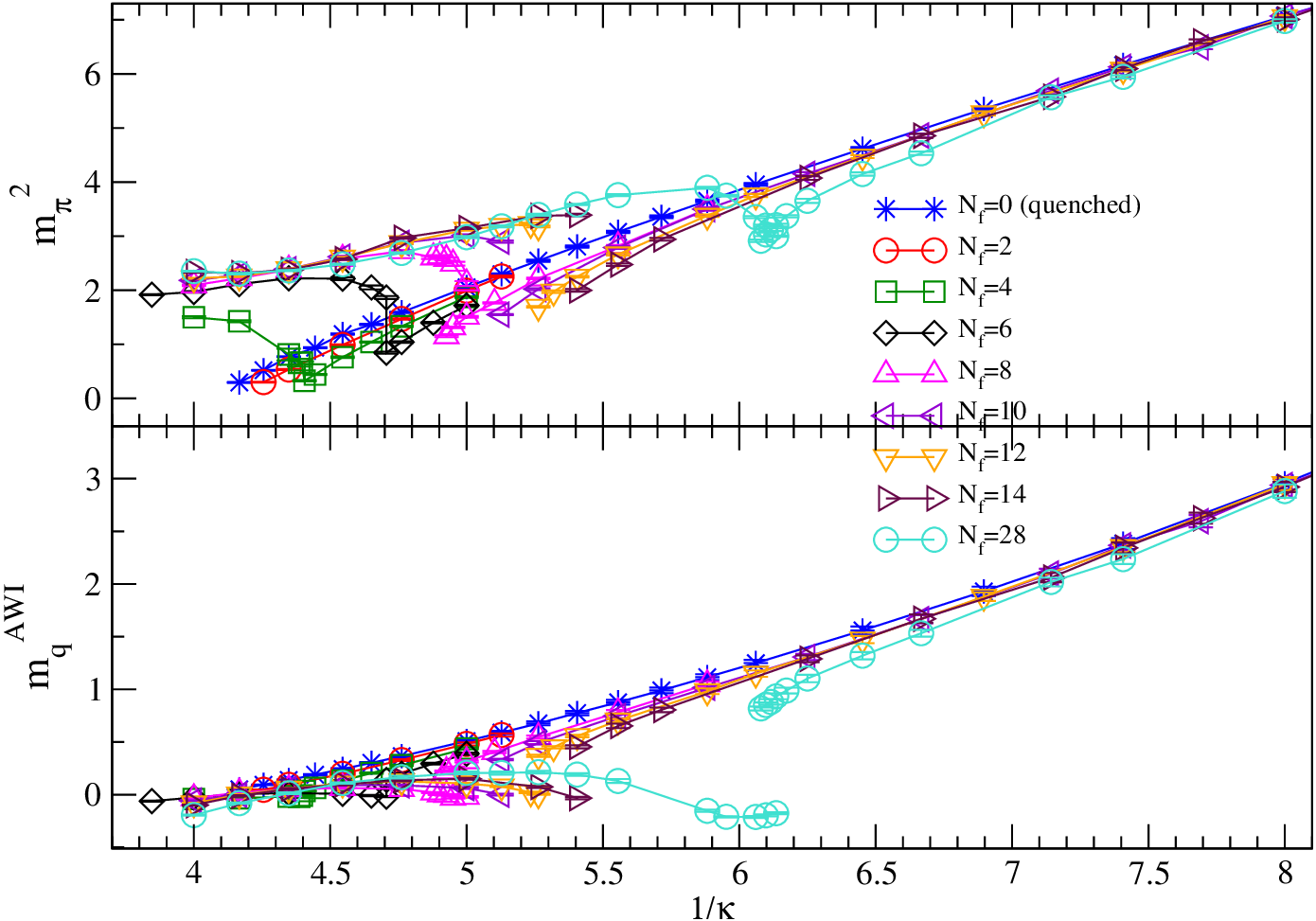}
\caption{
$m_\pi^2$ and $m_q^{\rm AWI}$ as
functions of
$1/\kappa$ in $SU(2)$ gauge theory with $N_f$ flavours at $\beta=0$.
}
\label{fig:mpib0}
\end{figure}

Fig.~\ref{fig:mpib0} shows $m_\pi^2$ and $m_q^{\rm AWI}$ 
over a wide range of $1/\kappa$.
Notice that
$m_\pi^2$ and $m_q^{\rm AWI}$ become insensitive to $N_f$ as $1/\kappa$
increases.  This is no surprise because sufficiently heavy quarks have a
minimal effect on the vacuum structure of the theory, so meson and quark
masses for any $N_f>0$ approach their quenched ($N_f=0$) values as $1/\kappa$
grows.

\begin{figure}[tbh]
\hspace{2cm}\includegraphics[width=12cm,trim=0 0 0 0,clip=true]{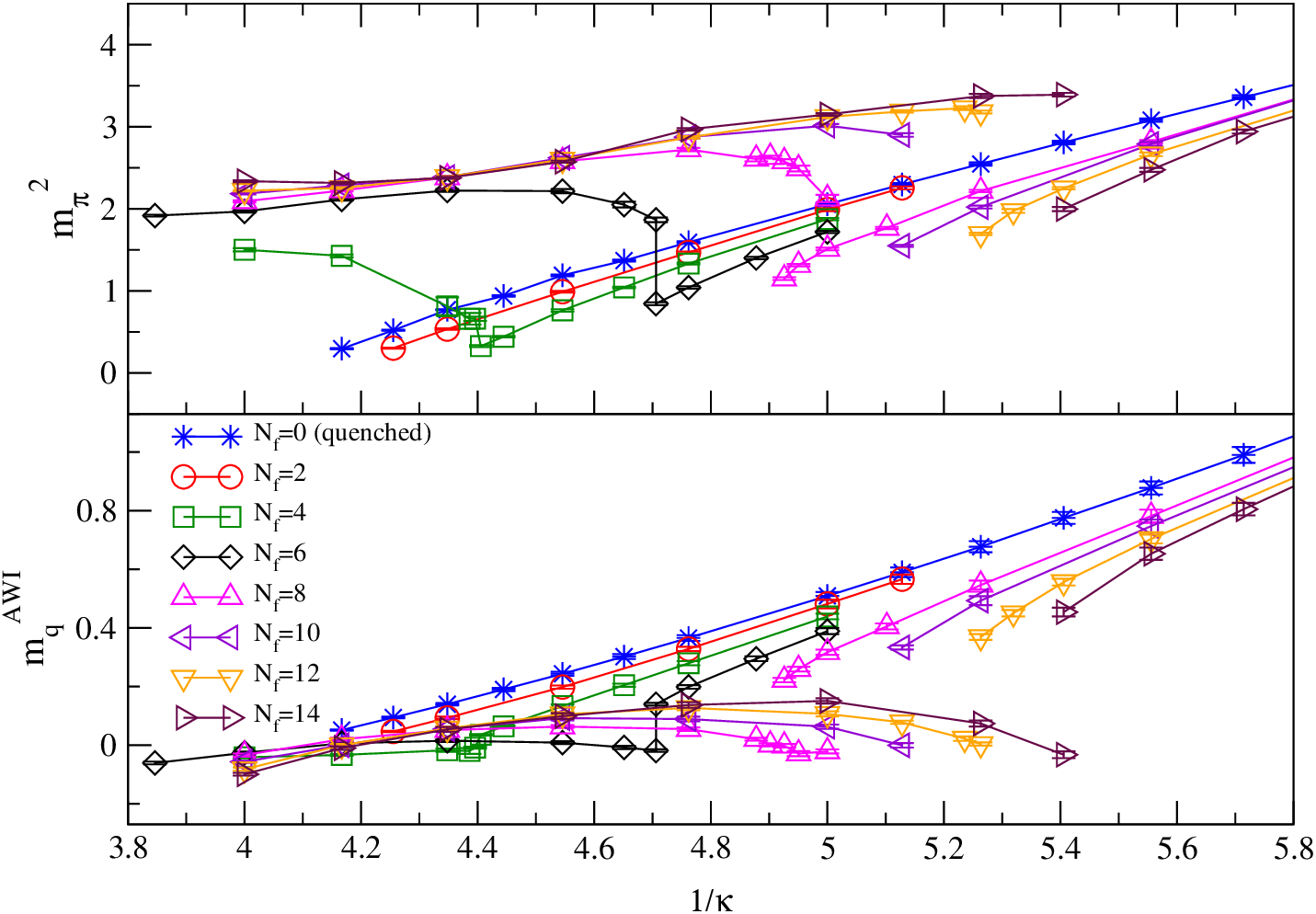}
\caption{
Same as Figure~\ref{fig:mpib0} but for a smaller range of $1/\kappa$.
}
\label{fig:mpib0big}
\end{figure}
A close-up of the small $1/\kappa$ region is displayed in
Fig.~\ref{fig:mpib0big}.  The same transition that was observed in the
elementary plaquette is clearly seen in $m_\pi^2$ and $m_q^{\rm AWI}$ as well.
{}From these data it is clear that the transition exists at least
for $N_f\geq4$ and the two-state signals imply that it is first order at least
for $N_f\geq6$.
The transition moves to larger $1/\kappa$ as $N_f$ is increased
and we speculate that the transition will approach $\kappa=\nicefrac{1}{8}$ as
$N_f\to\infty$, i.e.\
the gauge theory with infinitely many fermions will become the free theory.

\begin{figure}[tbh] 
\hspace{2cm}\includegraphics[width=12cm,trim=0 0 0 0,clip=true]{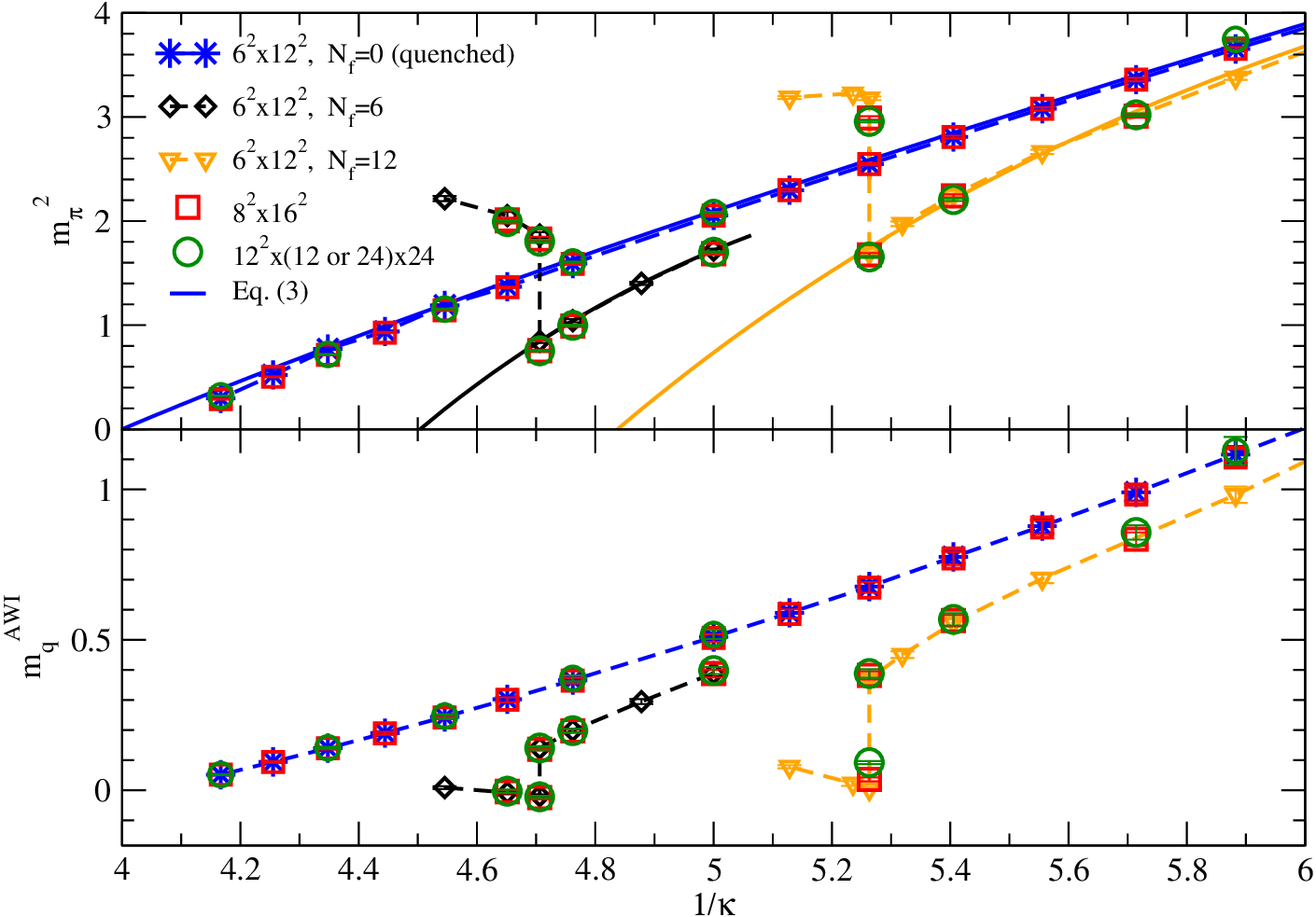}
\caption{
Volume independence of $m_{\pi}^2$ and $m_q^{\rm AWI}$ as a function of $1/\kappa$ 
in $SU(2)$ gauge theory for $N_f=6$  and 12 at $\beta=0$.
The prediction from Eq.~(\ref{eq:pion}) is shown as a solid blue curve
that reaches $m_\pi^2=0$ at $1/\kappa=4$.
Quadratic fits to lattice data for $m_\pi^2$ are shown as a pair of solid
curves ($N_f=6$ is black and $N_f=12$ is orange).
}
\label{fig:mpib0nf6}
\end{figure}
It should be noted that Fig.~\ref{fig:plaqnc2b0} is obtained for all four of
our lattice volumes: no volume dependence is observed.
This was expected in advance because the average plaquette is a short distance
quantity.  We have similarly confirmed volume independence for
$m_\pi^2$ and $m_q^{\rm AWI}$ over our range of volumes, and examples are
presented in Fig.~\ref{fig:mpib0nf6}.  In particular, the transition shows no
evidence of vanishing in the infinite volume limit, and it shows no evidence
of moving to $\kappa=\nicefrac{1}{4}$ in the infinite volume limit.

Besides the transition itself, 
Figs.~\ref{fig:mpib0big} and \ref{fig:mpib0nf6}
show an interesting curvature at values of $1/\kappa$ above the transition,
where the slope deviates more and more from the quenched slope when
approaching the transition point.
If someone wanted to extrapolate to $m_\pi^2=0$ using only data above the
transition, then the critical hopping parameter so obtained, let's name it
$\kappa_c^{\rm ext}$, would be smaller than $\nicefrac{1}{4}$.
Of particular interest is the fact that even the data above the transition
in Figs.~\ref{fig:mpib0big} and
\ref{fig:mpib0nf6} do not follow Eq.~(\ref{eq:pion}) and do not arrive at
$\kappa_c^{\rm ext}=\nicefrac{1}{4}$ despite Eq.~(\ref{eq:standardkappa}).

To ascertain the properties of the transition and of the regions it separates,
a range of other observables should be discussed.  That discussion goes beyond
the scope of the present work, but brief comments and selected plots can be
found in Appendix A.

\section{{\boldmath $SU(3)$} with many flavours at \boldmath{$\beta=0$}}

\begin{figure}[tbh] 
\hspace{2cm}\includegraphics[width=12cm,trim=0 0 0 0,clip=true]{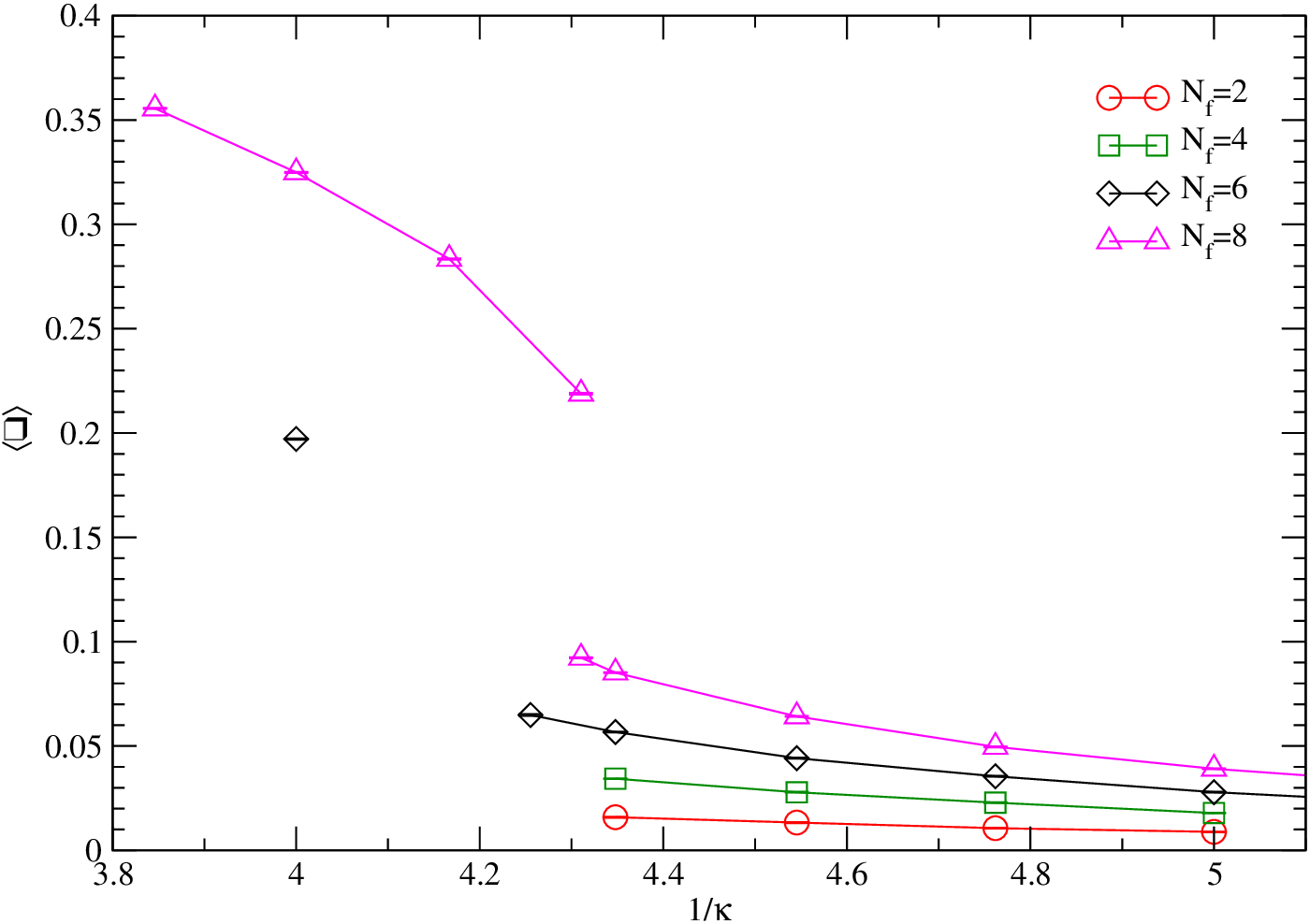}
\caption{
Average plaquette as a function of $1/\kappa$ 
in $SU(3)$ gauge theory with $N_f$ flavours at $\beta=0$.
}
\label{fig:plaqnc3b0}
\end{figure}

\begin{figure}[tbh] 
\hspace{2cm}\includegraphics[width=12cm,trim=0 0 0 0,clip=true]{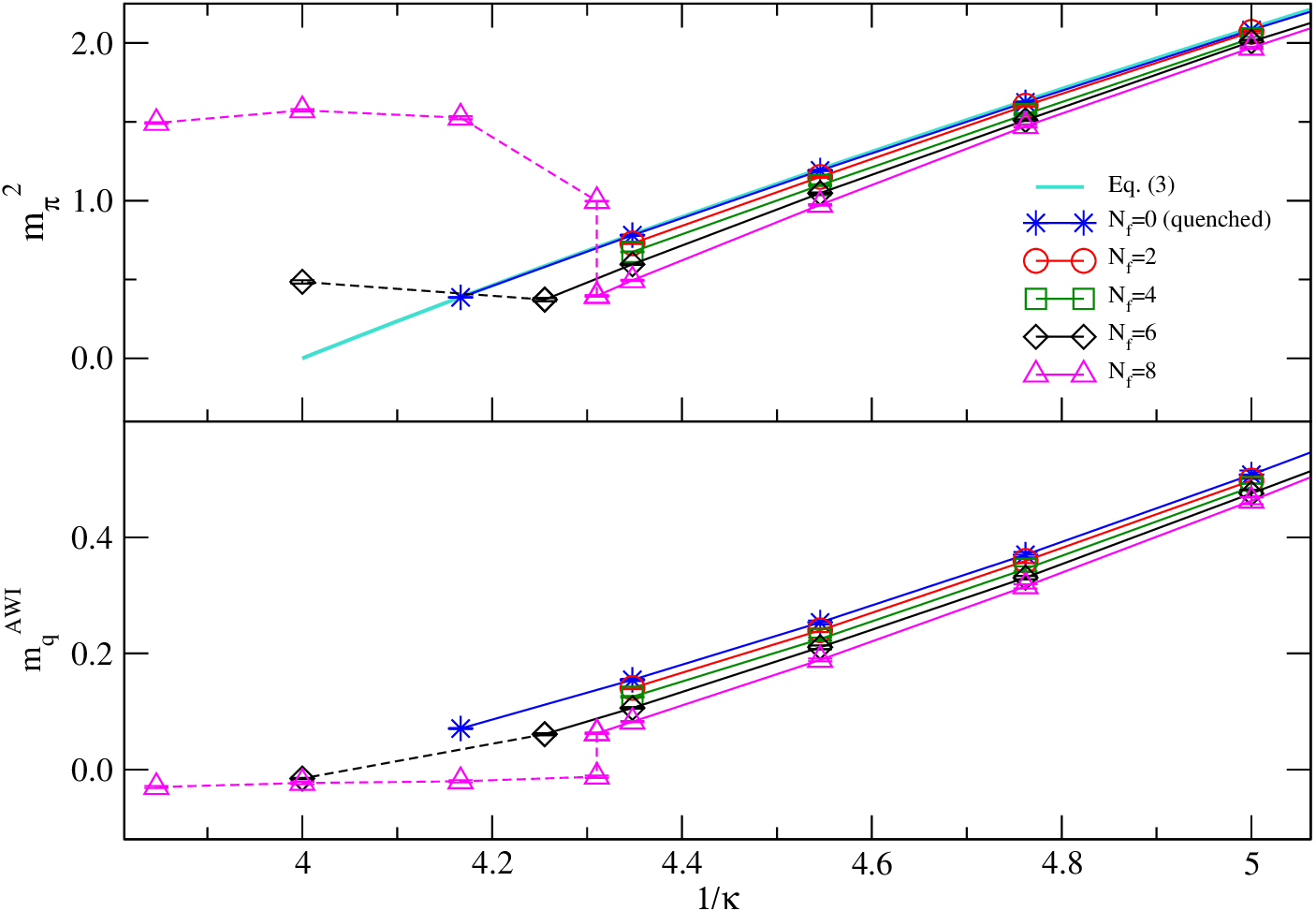}
\caption{
$m_\pi^2$ and $m_q^{\rm AWI}$ 
as functions of
$1/\kappa$ in $SU(3)$ gauge theory with $N_f$ flavours at $\beta=0$.
The prediction from Eq.~(\ref{eq:pion}) is shown as a solid blue curve
reaching $m_\pi^2=0$ at $1/\kappa=4$.
}
\label{fig:nc3mpib0}
\end{figure}

Qualitatively the situation for $SU(3)$ is similar to that for $SU(2)$, but our
quantitative results indicate that deviations from Eq.~(\ref{eq:pion}) are
smaller for $SU(3)$.  Perhaps a large-$N_c$ suppression mechanism is at work.
Data for the average
plaquette, pseudoscalar meson mass, and axial-Ward-Takahashi-identity quark mass
are plotted in Figs.~\ref{fig:plaqnc3b0} and \ref{fig:nc3mpib0}.
Notice that $m_\pi^2$ is consistent with Eq.~(\ref{eq:pion}) for $N_f=0$, but
the $N_f>0$ data display bending that is very similar to the $SU(2)$ case but
of smaller magnitude.

The search for a possible transition is more expensive for $SU(3)$ than for
$SU(2)$ because we must work at larger $\kappa$.
A transition is found to exist for
$N_f\geq8$ but we cannot say whether it is abrupt.  
Recall 
that Ref.~\cite{Iwasaki} found the transition to occur only for
$N_f\geq7$; they
rely to some extent on counting the conjugate
gradient iterations required during thermalization
to determine the phase for $N_f=6$ at $\kappa=1/4$.
(We comment briefly in Appendix B.)
Our results\footnote{The configurations of
$N_f=6$ at $\kappa=0.25$ in our case are generated 
from the last configuration of $N_f=8$ at $\kappa=0.25$.
We use the HMC algorithm with
a time step of $\Delta \tau=0.005$, 
while Ref.~\cite{Iwasaki} uses the approximate $R$
algorithm with $\Delta \tau=0.01$.}
do show hints of a transition
for $N_f=6$.

\section{Discussion}

Simulations on small space-time lattices in the strong coupling limit display
the phase transition already discussed in
Ref.~\cite{Iwasaki}, but the present work makes the additional observation
that the dependence of the pseudoscalar meson mass on the hopping parameter
deviates from the expected formula, Eq.~(\ref{eq:pion}), 
as seen in Figs.~\ref{fig:mpib0nf6} and \ref{fig:nc3mpib0}.  
These deviations, which appear
as bending in plots of meson (and quark) mass versus hopping parameter, are
easily seen for $N_c=2$ and are smaller but still observed for $N_c=3$.
The formula in Eq.~(\ref{eq:pion}) was derived through a large-$N_c$
expansion\cite{Aoki}, and
the present data for $N_c=2,3$ suggest a rather rapid approach to
Eq.~(\ref{eq:pion}) as $N_c$ is increased.
Eq.~(\ref{eq:pion}) is also obtained for any $N_c$ with $N_f=0$.

Because of the phase transition there is generally no value of the
hopping parameter at which the pseudoscalar meson becomes massless, but
extrapolation from above the phase transition (i.e.\ from smaller hopping
parameters) does allow the definition of an effective critical hopping
parameter, $\kappa_c^{\rm ext}(N_f)$.  This extrapolated value is found to
obey
\begin{equation}\label{ourshift}
\kappa_c^{\rm ext}(N_f) <
\kappa_c^{\rm ext}(N_f-1) <
\ldots <
\kappa_c^{\rm ext}(1) <
\kappa_c^{\rm ext}(0) = \frac{1}{4}
\end{equation}
as a consequence of the observed deviation away from Eq.~(\ref{eq:pion}).

The authors of Ref.~\cite{Iwasaki} performed a detailed study of
$SU(3)$ gauge theory with $N_f$ Wilson fermions 
and from that study they proposed
a phase structure with regions of confinement and deconfinement for
certain values of $N_f$.  They also include a brief appendix (in the most
recent publication in \cite{Iwasaki}) devoted to
the $SU(2)$ theory, where $\kappa_c=\nicefrac{1}{4}$ is assumed and simulations
at that hopping parameter are used as the basis for conclusions about a
confinement/deconfinement transition.  The results of
Ref.~\cite{Iwasaki} differ from those obtained using other
actions\cite{Fleming:2008gy}, and one might ask whether assumptions about
$\kappa_c=\nicefrac{1}{4}$ might play a role.  However, Eq.~(\ref{ourshift})
confirms that for any $N_f$
\begin{equation}
\kappa_d(N_f) < \kappa_c^{\rm ext}(N_f) \leq \frac{1}{4} \,,
\end{equation}
where $\kappa_d$ is the location of the phase transition;
this fact is sufficient
to leave the conclusions of Ref.~\cite{Iwasaki} intact.
Therefore the source of discrepancies between the phase structure obtained
from the Wilson action versus other actions must be sought elsewhere.

\begin{figure}[tbh]
\hspace{2cm}\includegraphics[width=12cm,trim=0 0 0 0,clip=true]{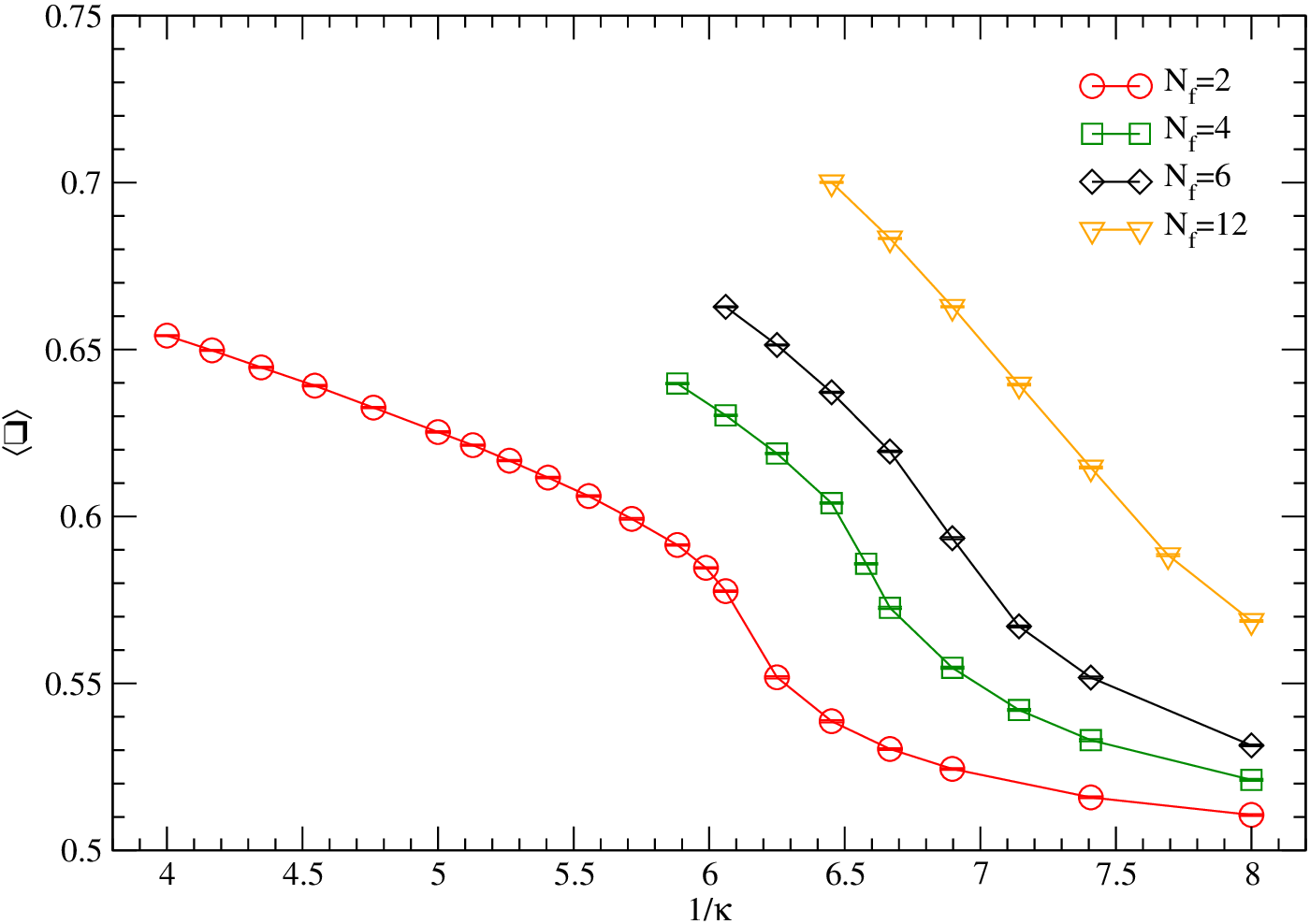}\caption{
Average plaquette as a function of $1/\kappa$
in $SU(2)$ gauge theory with $N_f$ flavours at $\beta=2$ on $8^2\times16^2$
lattices.
}
\label{fig:plaqb2}
\end{figure}

\begin{figure}[tbh]
\hspace{2cm}\includegraphics[width=12cm,trim=0 0 0 0,clip=true]{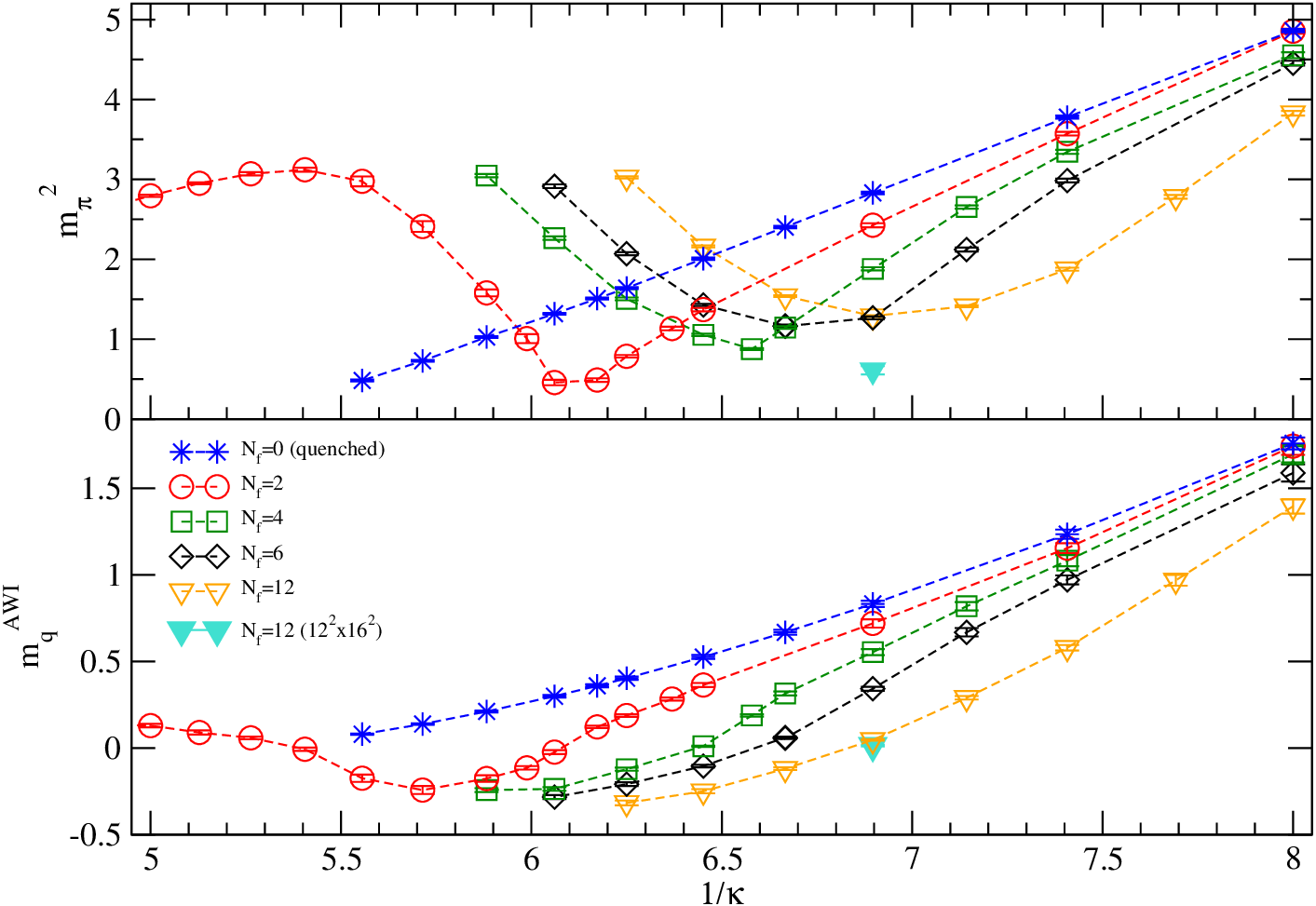}
\caption{
$m_\pi^2$ and $m_q^{\rm AWI}$ 
as functions of
$1/\kappa$ in $SU(2)$ gauge theory with $N_f$ flavours at $\beta=2$.
Lattices are $8^2\times16^2$ except for one point labelled as $12^2\times16^2$.
}
\label{fig:mpib2}
\end{figure}

It is worth noting that phenomena in the strong coupling limit are not typical
of results at weaker couplings.  For example the average plaquette,
meson mass, and quark mass for the $SU(2)$ theory at $\beta=2$ are plotted in
Figs.~\ref{fig:plaqb2} and \ref{fig:mpib2}.  Instead of an abrupt transition
that is essentially independent of lattice volume, we now see smoothly
continuous functions and the meson mass has a significant volume
dependence.  Detailed studies at $\beta\neq0$ are left for future work.

\section*{Appendix A: Additional observables}

\begin{figure}[tbh]
\hspace{2cm}\includegraphics[width=12cm,trim=0 0 0 0,clip=true]{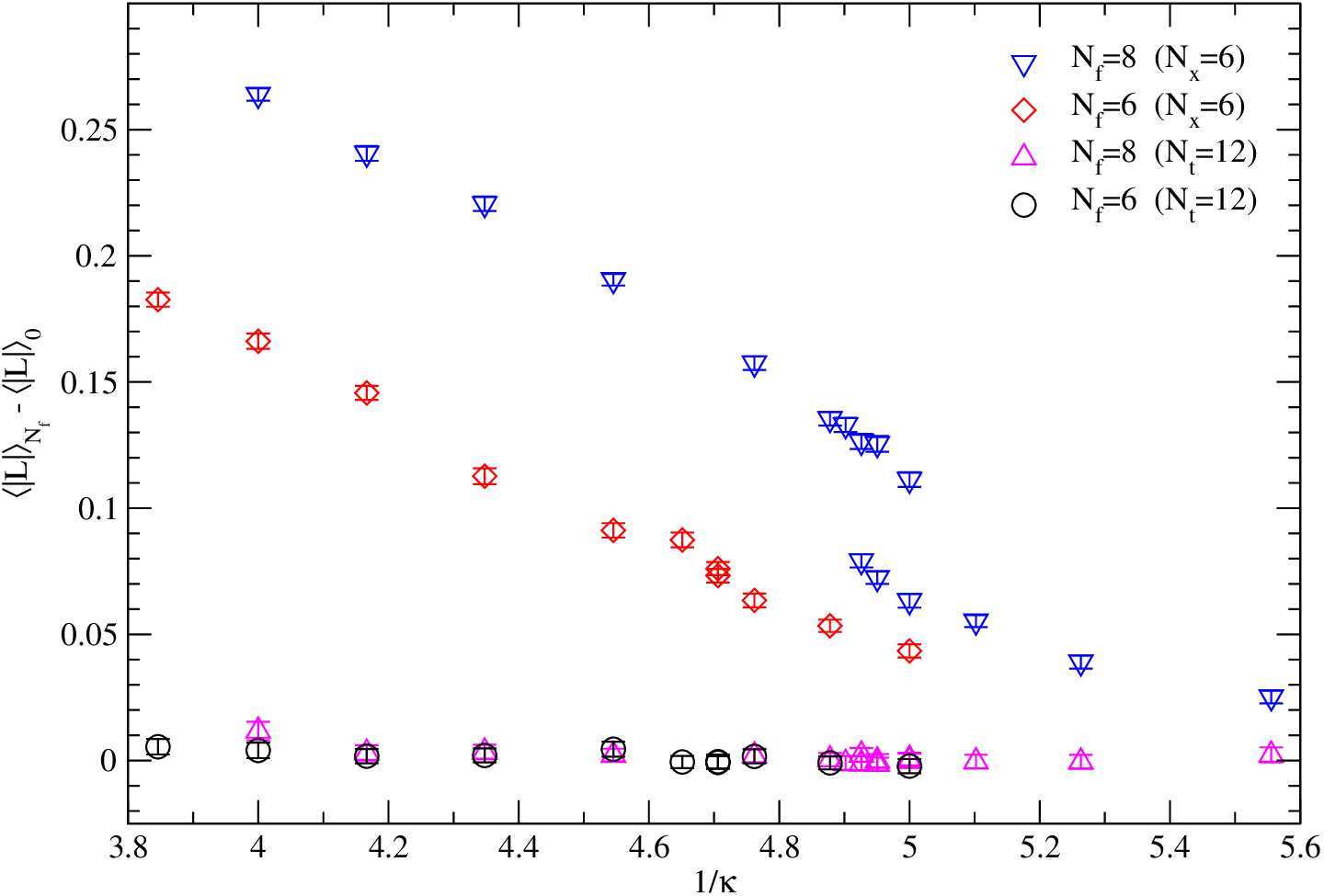}
\caption{
The dependence of a Polyakov loop $\langle| L|\rangle$
on $N_f$ flavours is expressed through the difference
$\langle| L|\rangle_{N_f} - \langle| L | \rangle_0$.
This difference is plotted as a function of $1/\kappa$ in $SU(2)$ gauge theory
on $6^2\times12^2$ lattices at $\beta=0$ with $N_f=6$ and 8.
}
\label{fig:polyakov}
\end{figure}

A Polyakov loop is the trace of a product of links in a straight line along a
lattice axis which, via periodic boundary conditions, closes upon itself.
Polyakov loops are often used as an indicator for confinement.
The absolute value of a Polyakov loop
is plotted in Fig.~\ref{fig:polyakov}
at $\beta=0$ on $6^2\times12^2$ lattices.
For each of the two cases $N_f=6$ and $N_f=8$, the graph shows the Polyakov
loops in both the shorter ($N_x=6$) and longer ($N_t=12$) lattice directions.
The shorter Polyakov loops are numerically large and dependent on the
hopping parameter, with a two-state signal at the transition for large
enough $N_f$.
The longer Polyakov loops are consistent with zero for all values of $\kappa$.
We refrain from drawing conclusions about confinement from this brief
consideration of Polyakov loops.

For discussions of chiral symmetry and its possible breaking, natural
quantities include the chiral condensate 
$\left< \psibar\psi\right>$
and the pseudoscalar density 
$\left< \psibar \gamma_5 \psi \right>$.
The low-lying eigenvalues of the Wilson-Dirac operator relate to 
these observables
through the spectral representation of the quark propagator.
For instance, $\left< \psibar \gamma_5 \psi \right> 
= \sum_{\lambda} \frac{\langle \lambda | \lambda \rangle}{\lambda(H_W)}$
and 
$\left< \psibar \psi \right> 
= \sum_{\lambda} \frac{\langle \lambda \left| \gamma_5 \right| \lambda \rangle}{\lambda(H_W)}$,
where $\lambda(H_W)$ and $\large| \lambda \rangle$ are
an eigenvalue and eigenvector respectively
of the hermitian Wilson-Dirac operator $H_W=\gamma_5 D_W$.

\begin{figure}[tbh]
\hspace{2cm}\includegraphics[width=12cm,trim=0 0 0 0,clip=true]{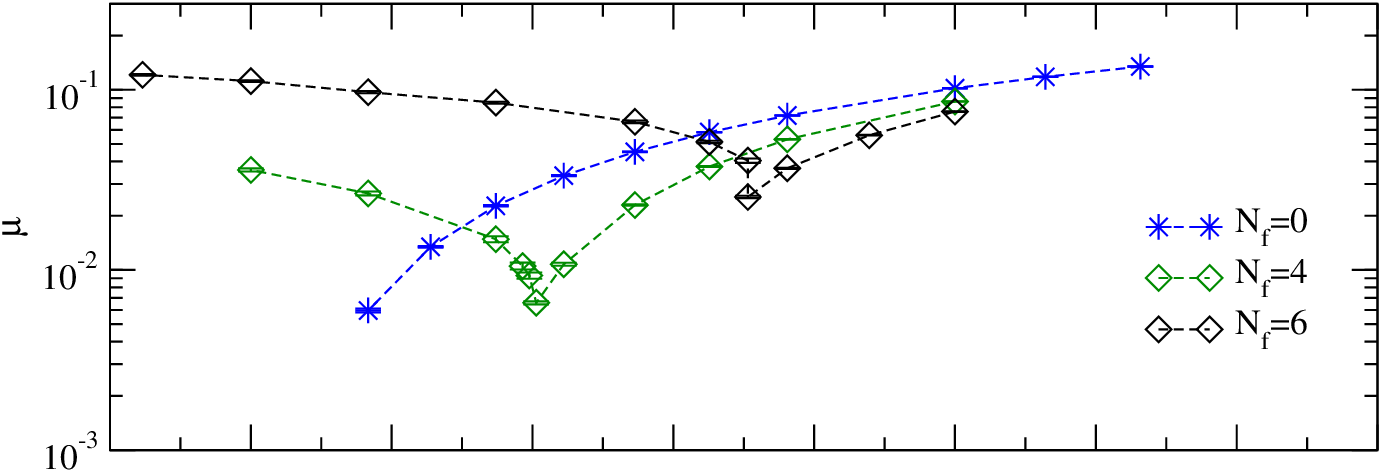}
\vspace{-0.7mm}

\hspace{21.5mm}\includegraphics[width=12.05cm,trim=0 0 0 0,clip=true]{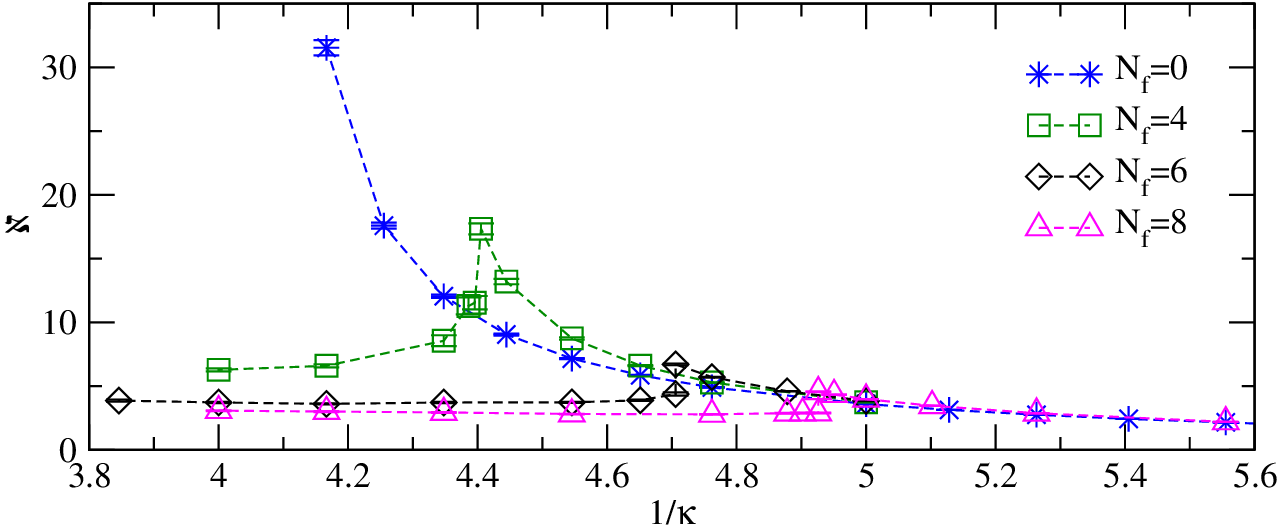}
\caption{
$SU(2)$ gauge theory with $N_f$ flavours 
at $\beta=0$ on a $6^2\times12^2$ lattice.
Upper panel: 
$\mu=\sqrt{\lambda_0(H_W^2)}$,
where $\lambda_0(H_W^2)$ is the lowest eigenvalue of $H_W^2$, 
as a function of $1/\kappa$.
Lower panel: 
${\cal N}=(2\kappa)^2\sum_{{\vec x},t}\langle P(\vec x,t)P(0)
\rangle$ as a function of $1/\kappa$.
}
\label{fig:propnorm}
\end{figure}

The upper panel of Fig.~\ref{fig:propnorm} is 
the lowest eigenvalue, $\mu$, 
defined by $\mu=\sqrt{\lambda_0(H_W^2)}$
in the eigenvalue equation, 
$H_W^2 | \lambda_0 \rangle = \lambda_0 | \lambda_0 \rangle$,
and obtained by the Ritz functional method.
This behaviour hints at the dependence
$\left< \psibar \gamma_5 \psi \right> \sim 1/\mu$.
At least for $N_f\geq4$, two phases are found
and in both phases the eigenvalues are not particularly small.

Because Wilson fermions break chiral symmetry explicitly, 
the chiral condensate is not adequate 
and should be replaced by the subtracted chiral condensate defined by
$(\psibar\psi)_{\rm subt} = 2m_q^{\rm AWI} {\cal N}$
where
${\cal N} = (2\kappa)^2\sum_{\vec x,t}\bigg<P(\vec x,t)P(\vec 0,0)\bigg>$.
The fermion mass $m_q^{\rm AWI}$ itself is discussed in the main body of this
article, and it is most naturally defined only for $1/\kappa$ values above
the transition, so here we focus on the observation of ${\cal N}$.
The lower panel of Fig.~\ref{fig:propnorm} displays $\cal N$ for
$SU(2)$ gauge theory at strong coupling for various $N_f$.
Indications of a $1/m_q$ divergence in the confinement phase (i.e.\ $1/\kappa$
above the transition) are visible, as expected.
For both observables in Fig.~\ref{fig:propnorm},
the transition is visible when $N_f\geq4$.

\section*{Appendix B: Algorithmic issues}

\begin{figure}[tbh]
\includegraphics[width=7.3cm,trim=0 0 0 0,clip=true]{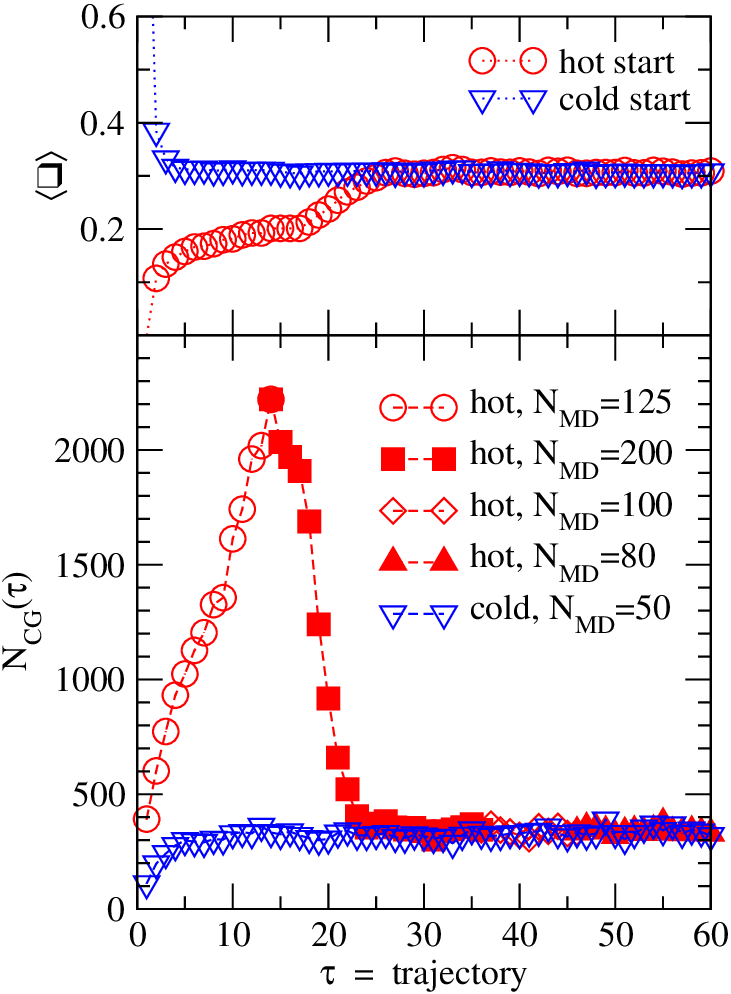}
\includegraphics[width=7.8cm,trim=0 0 0 0,clip=true]{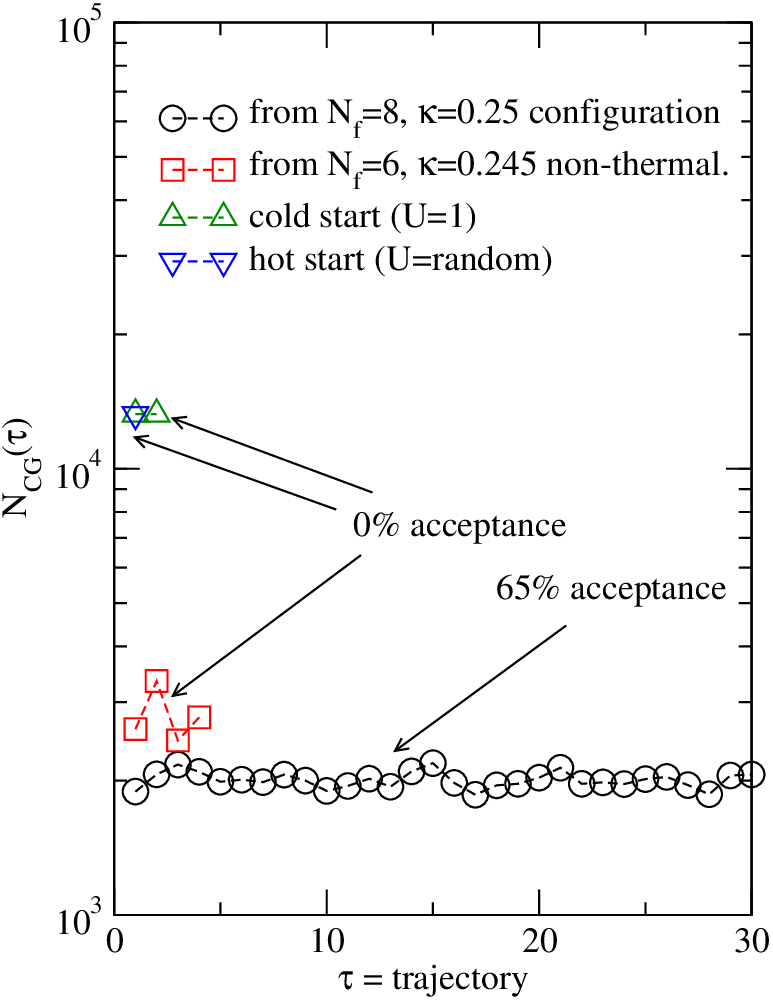}
\caption{
Left panel:
The average plaquette and $N_{CG}(\tau)$ versus trajectory index $\tau$
in $SU(2)$ gauge theory on $8^2\times16^2$ lattice 
at $\beta=0$ and $\kappa=0.215$ with $N_f=6$.
Right panel:
$N_{CG}(\tau)$ versus $\tau$
in $SU(3)$ gauge theory  with $N_f=6$ on $6^2\times12^2$ lattice 
at $\beta=0$, $\kappa=1/4$ and $N_{MD}=200$.
}
\label{fig:cgsu2}
\end{figure}

The HMC algorithm uses a number of trajectories, traditionally distinguished
by an integer $\tau$.
Each trajectory uses some number of molecular dynamics steps, $N_{\rm MD}$,
and each of those uses some number of Conjugate Gradient (CG) iterations,
$n_i^{\rm CG}(\tau)$ with $1\leq i\leq N_{\rm MD}$, to attain the imposed
precision.
The average number of CG iterations per MD step is therefore
$N_{\rm CG}(\tau)=\frac{1}{N_{\rm MD}}\sum_{i=1}^{N_{\rm MD}}n_i^{\rm CG}(\tau)$.
The left panel of Fig.~\ref{fig:cgsu2} shows this quantity for sequential trajectories during
thermalization for $SU(2)$ with $N_f=6$, $\beta=0$, and $\kappa=0.215$
on a $8^2\times16^2$ lattice with precision defined by
$||r_{\rm CG}^{\rm res}||<10^{-12}$.
For reference, the average plaquette is also plotted for each trajectory.
The hot start requires more trajectories before reaching thermalization,
and some of those trajectories needed many CG iterations.
One could possibly be tempted to abort such a calculation before
reaching equilibrium and assume that the increasing $N_{\rm CG}(\tau)$ would
never return to a smaller value at larger $\tau$.
The right panel of Fig.~\ref{fig:cgsu2} shows $N_{CG}(\tau)$
for $SU(3)$ at $N_f=6$, $\beta=0$, and $\kappa=0.250$
on a $6^2\times12^2$ lattice with $||r_{\rm CG}^{\rm res}||<10^{-12}$
and $N_{MD}=200$.
When configuration generation starts from
the last configuration of $N_f=8$ at $\kappa=0.250$,
the HMC becomes stable 
at $65\%$ acceptance and $N_{CG}(\tau)=O(10^3)$.
On the other hand,
the hot and cold starts are not accepted by the HMC's Metropolis test
with $N_{CG}(\tau)=O(10^4)$.
In both cases\footnote{We plotted the beginning $O(10)$ trajectories 
in about 300 runs.} 
of Fig.~\ref{fig:cgsu2},
the large value of $N_{CG}(\tau)$ does not imply the existence
of a massless mode.

\section*{Acknowledgments}

This work was supported in part by the Natural Sciences and Engineering
Research Council of Canada, and was made possible by the facilities of the
Shared Hierarchical Academic Research Network (SHARCNET:www.sharcnet.ca).
G.K.\ thanks the University of Regina for hospitality during her stay there,
and we acknowledge use of Regina's VXRACK computer cluster.


\end{document}